\documentstyle[preprint,aps]{revtex}
\tightenlines 

\begin{document}

\draft
\title{Integral equation formulation of the spinless Salpeter equation}
\author{F. Brau\thanks{Chercheur I.I.S.N.}\thanks{E-mail: 
fabian.brau@umh.ac.be}} 
\address{Universit\'{e} de Mons-Hainaut, Place du Parc 20,
B-7000 Mons, BELGIQUE}
\date{September 1997}

\maketitle

\begin{abstract}
The spinless Salpeter equation presents a rather particular differential 
operator. In this paper we rewrite this equation into integral and  
integro-differential equations. This kind of equations are well known and 
can be more easily handled. We also present some analytical results 
concerning the spinless Salpeter equation and the action of the 
square-root operator.
\end{abstract}

\section{Introduction}
\label{sec:intro}

The Schr\"{o}dinger equation is a very well defined partial derivative 
equation since its differential operator is a Laplacian. Moreover it 
reduces to a simple differential equation when the interaction is central. 
This equation has been intensively studied and it is well understood. A 
simple relativistic version of the Schr\"{o}dinger equation, the Spinless 
Salpeter Equation (SSE), presents more difficulties. This equation is,
\begin{equation}
\label{eq1}
\left(\sqrt{\vec p\,^2 + m_1^2} + \sqrt{\vec p\,^2 + m_2^2}\right) 
\Psi(\vec r\,) = \Bigl(E - V(\vec r\,)\Bigr)\, \Psi(\vec r\,),
\end{equation}
where $m_1$, $m_2$ are the masses of the particles, $\vec p$ is their 
relative momentum, $V(\vec r\,)$ is the potential interaction and $E$  
the eigenenergy of the stationary state $\Psi(\vec r\,)$ ($\hbar=c=1$). 
$\vec p$ and $\vec r$ are conjugate variables. Actually, this last 
equation is not so well defined since the kinetic energy operator is a 
nonlocal one. Its action on a function $f(\vec r\,)$ is known only if 
$f(\vec r\,)$ is an eigenfunction of the operator $\vec p\,^2$. In this 
case we obtain,
\begin{equation}
\label{eq2}
\sqrt{\vec p\,^2 + m^2}\ f(\vec r\,) = \sqrt{\alpha + m^2}\ f(\vec r\,),
\end{equation}
where $\alpha$ is the corresponding eigenvalue of $\vec p\,^2$. 
Consequently this operator is difficult to handle. However the SSE is not 
a marginal equation. The correct description of the bound states of two 
particles is achieved with the Bethe-Salpeter equation~\cite{salp51}. This 
last reduces to the SSE~\cite[p. 297]{grei94} when:  
\begin{itemize}
\item The elimination of any dependences on timelike variables is performed 
(which leads to the Salpeter equation~\cite{salp52}), and
\item Any references to the spin degrees of freedom of particles are 
neglected as well as negative energy solutions.
\end{itemize}

Despite the presence of a so particular operator, the SSE is often used 
(see for instance 
Refs.~\cite{stan80,cea83,gupt85,godf85,laco89,fulc94,sema97}) 
since its numerical resolution is very easy for bound states 
(see for example Refs.~\cite{fulc94,nick84,dura90,brau97}).
However, it would be interesting to reformulate this equation in a way 
to better understand it or even to better use it. The main problem is the 
nonlocality hidden inside the kinetic energy operator. The idea is thus to 
extract this nonlocality to put it into evidence. That is why in this work 
we rewrite the SSE as integral or integro-differential equations where the 
nonlocality is then explicit. Moreover these kinds of equations are well 
known and defined. Another integral equation, different from those 
presented here, was found in Ref.~\cite{nick84}. This formalism allows to 
define the action of the square-root operator on wave functions and 
calculate the resulting functions. 

This paper is organized as follows. In Sec.~\ref{sec:theory}, we present 
the integral and integro-differential equations and we calculate 
their kernels. In Sec.~\ref{sec:analsol}, we find some analytical results 
concerning the action of the square-root operator and solutions of the SSE.  
At last, we present, in Sec.~\ref{sec:summary}, a summary of this work. 

\section{Integral equation formulation}
\label{sec:theory}

It is possible to rewrite the SSE into different forms. The first one 
that we present is an integral equation. It is obtained using the Green 
functions of the kinetic energy operator. The corresponding kernels are 
different according to the particles are identical or not. The second one 
is an integro-differential equation. It is obtained taking the square of 
the square-root operator and is only valid for two identical particles. 
For these formulations of the SSE only local interactions can be used 
contrary to the formulation proposed in Ref.~\cite{nick84}.

\subsection{First form}
\label{subsec:firstf}
In the introduction we gave the operator expression of the SSE. We are 
going now to formulate the integral expression of this equation. To obtain 
it, we use the Green functions of the kinetic energy operator. We first 
present the equal masses case. Let us consider the following function,
\begin{equation}
\label{eq16}
G(\Delta) = \frac{1}{(2\pi)^3} \int \frac{\exp \left(-i \vec p.\vec \Delta 
\right)}{2 \sqrt{p^2 + m^2}}\, d\vec p ,
\end{equation}
with $\vec \Delta = \vec r - \vec r\,'$ and $\Delta = | \vec \Delta |$. 
Then, $G(\Delta)$ is the Green function of the equal masses square-root 
operator since,
\begin{equation}
\label{eq17}
2 \sqrt{\vec p\,^2 + m^2}\ G(\Delta) = \frac{1}{(2\pi)^3} \int 
\exp \left(-i \vec p.\vec \Delta \right)\, d\vec p = \delta^3(\vec r - 
\vec r\,').
\end{equation}
The solution of the SSE can be written as,
\begin{equation}
\label{eq18}
\Psi(\vec r\,) = \Psi_0(\vec r\,) + \int G(\Delta)\, \Bigl(E-V(\vec r\,')
\Bigr)\, \Psi(\vec r\,')\, d\vec r\,',
\end{equation}
with $\sqrt{\vec p\,^2 + m^2}\ \Psi_0(\vec r\,) = 0$.
The solutions of this last equation are $\Psi_0(\vec r\,) = 0$ and 
$\Psi_0(\vec r\,)=i_l(mr)\, Y_{lm}(\hat r\,)$, where $i_l(x)=\sqrt{\frac
{\pi}{2x}}\, I_{l+\frac{1}{2}}(x)$, $I_{\nu}(x)$ is a modified 
Bessel function~\cite[p. 952]{grad80}, $Y_{lm}(\hat r)$ is a spherical 
harmonic, $r=|\vec r\, |$ and $\hat r=\vec r/r$. In the 
appendix~\ref{appendb} we show that $\Psi_0(\vec r\,)=k_l(mr)\, 
Y_{lm}(\hat r\,)$ is not a solution, the function $k_l(x)$ being defined 
as equal to $\sqrt{\frac{2}{\pi x}}\, K_{l+\frac{1}{2}}(x)$, with 
$K_{\nu}(x)$ a modified Bessel function~\cite[p. 952]{grad80}. Thus only 
the vanishing solution is relevant for physical problems. One sees 
immediately that the utilization of nonlocal potentials is impossible if 
we want to obtain a 1-dimensional integral equation. Calculating the 
3-dimensional integral (\ref{eq16}) we obtain
\begin{equation}
\label{eq19}
G(\Delta) = \frac{m}{4\pi^2 \Delta}\, K_1(m\Delta).
\end{equation}

If we consider only the case of central potentials, the wave function can 
be written in the form $\Psi(\vec r\,)=R_l(r)\, Y_{lm}(\hat r)$ where 
$R_l(r)$ is the radial part of the wave function. If, moreover, we 
place the $z$ axis along $\vec r$, the angular dependence of $\Delta$ is 
just given by the angle between the vector $\vec r\,'$ and the $z$ axis. 
Then knowing that~\cite[p. 158]{vars88}
\begin{equation}
\label{eq8}
Y_{lm}(0,\varphi) = \sqrt{\frac{2 l + 1}{4 \pi}}\ \delta_{m0},
\end{equation}
Eq.~(\ref{eq18}) is written as 
\begin{eqnarray}
\label{eq9} 
\nonumber
R_l(r)\, \sqrt{\frac{2l+1}{4 \pi}}\, \delta_{m0} =&& \frac{m}{4 \pi^2}\, 
\int_{0}^{\infty} r'^2\, \Bigl(E-V(r')\Bigr)\, R_l(r')\, dr'\, 
\int_{0}^{\pi} \sin \theta'\, \frac{K_1(m\Delta)}{\Delta}\, d\theta' \\ 
&&\int_{0}^{2\pi} Y_{lm}(\hat{r}')\, d\varphi'. 
\end{eqnarray}
We can now perform the integration over the angular variables.
We have the well known relations 
\begin{equation}
\label{eq10}
\int_0^{2\pi} Y_{lm}(\hat r')\, d\varphi'= 2\pi \, Y_{lm}(\hat r')\ 
\delta_{m0},
\end{equation}
and~\cite[p. 133]{vars88},
\begin{equation}
\label{eq11}
Y_{lm}(\theta,\varphi) = \exp(im\varphi)\, \sqrt{\frac{2l+1 \ (l-m)!}
{4\pi \ (l+m)!}} \ P_l^m \left(\cos\theta\right),
\end{equation}
where the functions $P_l^m(x)$ are the Legendre 
functions~\cite[p. 998]{grad80}.
Finally, using Eqs.~(\ref{eq9}-\ref{eq11}) and the regularized function 
$u_l(r) = r R_l(r)$, we can write the spinless Salpeter equation on the 
form of the following integral equation:
\begin{equation}
\label{eq20}
u_l(r) =\frac{1}{2\pi} \int_0^\infty {\cal G}_l(mr,mr')\, \Bigl(E - V(r')
\Bigr)\, u_l(r')\, dr',
\end{equation}
with
\begin{equation}
\label{eq21}
{\cal G}_l(mr,mr') = m\, r\, r'\ \int_0^\pi 
\sin\theta'\, \frac{K_1(m \Delta)}{\Delta}\, P_l \left(\cos\theta'
\right)\, d\theta',
\end{equation}
where the functions $P_l(x)$ are the Legendre polynomials 
\cite[p. 1025]{grad80}. The functions ${\cal G}_l(mr,mr')$ are analytic for 
each value of $l$. Performing the transformation $y=\sqrt{r^2+r'\,^2-2\, 
r\, r' \cos\theta'}$, we obtain
\begin{equation}
\label{eq31}
{\cal G}_l(mr,mr')=m \int_{|r-r'|}^{r+r'} K_1(my)\, 
P_l \left(\frac{r^2+r'\,^2-y^2}{2\, r\, r'} \right)\, dy.
\end{equation}
For each value of $l$, we can find the primitive and thus calculate the 
kernel. Indeed, we have the following relations~\cite[p. 87]{magn66}:
\begin{mathletters}
\label{eq29}
\begin{eqnarray}
\label{eq29:a}
\int x^{\nu + 1}\, K_{\nu}(x)\, dx &=& -\, x^{\nu + 1}\, K_{\nu + 1}(x), \\
\label{eq29:b}
\int x^{-\nu + 1}\, K_{\nu}(x)\, dx &=& -\, x^{-\nu + 1}\, K_{\nu - 1}(x).
\end{eqnarray}
\end{mathletters}
With these ones it is easy to show that,
\begin{eqnarray}
\label{eq30}
\nonumber
\int x^\nu \, K_\mu(x)\, dx &=& -\, x^\nu\, K_{\mu - 1}(x)-(\nu+\mu-1)\, 
x^{\nu-1}\, K_\mu(x) \\ 
&+&(\nu+\mu-1)\, (\nu-\mu-1)\, \int x^{\nu-2}\, K_\mu(x)\, dx.
\end{eqnarray}
This last relation is useful when one calculates the kernel for $l \geq 2$. 
The modified Bessel functions recursion relation is~\cite[p. 970]{grad80}
\begin{equation}
\label{eq34}
K_{\nu + 1}(x)=\frac{2\, \nu}{x}\, K_\nu(x) + K_{\nu -1}(x).
\end{equation}
With Eq.~(\ref{eq31}) we see that the polynomial under the 
integral is composed only of even power of the integration variable. Then 
with relation~(\ref{eq30}), we can calculate analytically the kernel 
for each value of $l$.  
Another way (more simple) to calculate the kernel is to use the relation 
derived in Ref.~\cite{nick84}
\begin{equation}
\label{nick1}
{\cal G}_l(x,x')=2^l z^{l+1} \left(\frac{1}{z} \frac{\partial}
{\partial z}\right)^l \frac{1}{z} \Bigl[(y-z)^{\frac{l}{2}}\, 
K_l\left(\sqrt{y-z}\right)-(y+z)^{\frac{l}{2}}\, K_l
\left(\sqrt{y+z}\right)\Bigr],
\end{equation}
with $y=x^2+x'^2, z=2xx'$.
The kernel have as expression for $l=0$ and $l=1$:
\begin{equation}
\label{eq32}
{\cal G}_0(x,x')=K_0(|x-x'|)-K_0(x+x'),
\end{equation} 
\begin{eqnarray}
\label{eq33}
\nonumber 
{\cal G}_1(x,x')&=& K_0(|x-x'|)+K_0(x+x') \\ 
&+& \frac{1}{x x'} \Bigr[(x+x')\, K_1(x+x')-|x-x'|\, K_1(|x-x'|)\Bigl].
\end{eqnarray} 
The expressions (\ref{eq32}) and (\ref{eq33}) show that kernels decrease 
exponentially when arguments are important enough (($x+x')$ and 
$|x-x'| \gg 1$) since $K_\nu(x) \sim \sqrt{2/(\pi\, x)}\exp(-x)$ when 
$|x| \gg 1$~\cite[p. 961]{grad80}. This behavior indicates that when the 
masses of particles are large, the nonlocality becomes negligible, the SSE 
becomes almost local and its spectrum coincide with the Schr\"odinger 
spectrum. However the explicit limit cannot be done in this formalism. One 
can remark that these kernels present a logarithmic singularity for $r=r'$.

We treat now the unequal masses case. After the previous calculations, 
it is obvious that the Green function for the unequal masses kinetic energy 
operator is given by,
\begin{equation}
\label{eq23}
\tilde{G}(\Delta) = \frac{1}{(2\pi)^3} \int \frac{\exp \left(-i \vec p.\vec 
\Delta \right)}{\sqrt{p^2 + m_1^2}+\sqrt{p^2 + m_2^2}}\, d\vec p.
\end{equation}
This equation becomes, if $m_1 \neq m_2$,
\begin{equation}
\label{eq24}
\tilde{G}(\Delta) = \frac{1}{(2\pi)^3}\, \frac{1}{m_1^2 - m_2^2} 
\int \left(\sqrt{p^2 + m_1^2}- \sqrt{p^2 + m_2^2}\right)\, \exp \left(
-i \vec p.\vec \Delta \right)\, d\vec p.
\end{equation}
The solution of the SSE can be written as a similar form than 
Eq.~(\ref{eq18}) and the function $\Psi_0(\vec r\,)$ is here always null. 
One can find it easily by acting the operator 
$\left(\sqrt{\vec p\,^2 + m_1^2}- \sqrt{\vec p\,^2 + m_2^2}\right)$ 
on the equation $\left(\sqrt{\vec p\,^2 + m_1^2}+ 
\sqrt{\vec p\,^2 + m_2^2}\right)\, \Psi_0(\vec r\,)=0$.
Extracting the operator $\left(-\Delta + m^2\right)$ from Eq.~(\ref{eq24}), 
performing the integration over the momenta and using the Green's theorem 
(see Ref.~\cite{nick84}), one can show that
\begin{equation}
\label{new1}
\Psi(\vec r\,)=\frac{m_1}{2\pi^2\left(m_1^2-m_2^2\right)}\, \int 
\frac{K_1(m_1\Delta)}{\Delta}\left(-\Delta_{r'} + m_1^2\right)\Bigl(E-V(r')
\Bigr)\, \Psi(\vec{r}\,')\, d\vec{r}\, ' +(m_1 \rightarrow m_2).
\end{equation}
Again, placing the $z$ axis along $\vec r$ and repeating the previous 
calculation we obtain
\begin{equation}
\label{new2}
u_l(r)=\frac{1}{\pi \left(m_1^2-m_2^2\right)} \int_{0}^{\infty} 
\tilde{{\cal G}}_l(m_1r,m_1r')\Bigl(E-V(r')\Bigr)\, u_l(r')\, dr'
+(m_1 \rightarrow m_2),
\end{equation}
with
\begin{equation}
\label{new3}
\tilde{{\cal G}}_l(mr,mr')={\cal G}_l(mr,mr')\left[-\frac{d^2}{dr'^2}+
\frac{l(l+1)}{r'^2}+m^2\right].
\end{equation}

\subsection{Second form}
\label{subsec:secondf}
This second rewriting of the SSE leads to an integro-differential equation. 
By acting the kinetic energy operator on the left of the equal masses 
SSE we obtain,
\begin{equation}
\label{eq26}
4 \left(\vec p\,^2 + m^2\right)\, \Psi(\vec r\,) = E\, \Bigl(E-V(\vec r\,)
\Bigr)\,  \Psi(\vec r\,) - 2\, \sqrt{\vec p\,^2 + m^2}\ V(\vec r\,)\, 
\Psi(\vec r\,).
\end{equation}
Thus, we must calculate the action of the square-root operator on the 
product of functions $V(\vec r\,)\, \Psi(\vec r\,)$. If we consider only 
central potentials this product gives a function for which the angular 
dependence is a spherical harmonic. One can show, with the formalism 
developed in Ref.~\cite{nick84}, that the radial part $g(r)$ of the 
function $g(r)Y_{lm}(\hat r)$, resulting from the action of the kinetic 
energy operator on the function $h(r)Y_{lm}(\hat r)$, is given by
\begin{equation}
\label{new4}
g(r)=\frac{1}{\pi r}\int_{0}^{\infty} \tilde{{\cal G}}_l(mr,mr')\, 
\tilde{h}(r')\, dr',
\end{equation}
with $\tilde{h}(x)=x\, h(x)$. When the right-hand side integral is relevant, 
this rewriting gives meaning to the action of the kinetic energy operator 
on a central problem wave function. With this result, one obtain from 
Eq.~(\ref{eq26})
\begin{equation}
\label{eq27}
\left[\frac{d^2}{dr^2}-\frac{l(l+1)}{r^2}-m^2+\frac{E\Bigl(E-V(r)\Bigr)}
{4}\right]\, u_l(r) = \frac{1}{2\pi} \int_0^\infty 
\tilde{{\cal G}}_l(mr,mr')\, V(r')\, u_l(r')\, dr'.
\end{equation} 
This last equation is interesting because this is a kind of nonlocal 
Klein-Gordon Equation (KGE). This shows how the SSE differs from KGE when 
a nonvanishing interaction is introduced (the free SSE leads to a the free 
KGE with a different wave number $k$). Thus the difference is partially 
given by the nonlocal right-hand side of Eq.~(\ref{eq27}). 

To show how the square-root operator can be particular, we give a simple 
example in which we show how the hidden nonlocality of the kinetic energy 
operator could lead to some problems. Let us consider a square well with a 
range $r_c$ and a depth $-V_0$. In this case, according to $r$ is less 
than $r_c$ or not, the right-hand side of equal masses SSE is a constant 
($E$ or $E+V_0$) times the wave function. Then one can think that in each 
area this equation leads to,
\begin{equation}
\label{eq28}
\left[\Delta + \text{Sgn}(r_c-r)\, k^2\right]\, \Psi(\vec r\,) = 0 ,
\end{equation}
with
\begin{eqnarray*}
k^2 &=& \frac{(E+V_0)^2}{4}-m^2 \quad \text{if} \quad r<r_c, \\
k^2 &=& m^2-\frac{E^2}{4} \quad \text{if} \quad r>r_c. 
\end{eqnarray*}
The continuity conditions, for the interior and exterior solutions, at 
distance $r=r_c$, fix the wave function and the 
energy. But this method to resolve the SSE does not take into account the 
nonlocality of the kinetic operator and we obtain the same kind of solution 
as the KGE. Actually, the problem is the discontinuity. The potential is 
not a constant for all values of $r$ and the kinetic operator does not 
commute with it. Thus the equal masses SSE does not lead to 
Eq.~(\ref{eq28}). However, in Eq.~(\ref{eq27}) the nonlocality is 
contained in the potential part and we can now write an equation for each 
area ($r<r_c$ and $r>r_c$). We remark that this equation does not reduce to 
the free KGE when $r>r_c$ because the right-hand side does not vanish. Thus 
Eq.~(\ref{eq27}) leads to different solutions that these obtained with KGE 
for the same square well.

We conclude this section with some remarks concerning these integral 
equations. The equation derived in Ref.~\cite{nick84} is, for equal masses 
case, 
\begin{equation}
\label{new5}
\Bigl[E-V(r)\Bigr]\, u_l(r)=\frac{2}{\pi} \int_{0}^{\infty}
\tilde{{\cal G}}_l(mr,mr') \, u_l(r')\, dr'.
\end{equation}
Thus if the potential is just a constant, $V(r)=V$, using this equation and 
Eq.~(\ref{eq27}) we obtain,
\begin{equation}
\label{eq56}
\left[\frac{d^2}{dr^2}-\frac{l(l+1)}{r^2}-m^2+\frac{(E-V)^2}{4}\right]\, 
u_l(r)=0.
\end{equation}
The nonlocality disappears and we find the corresponding Klein-Gordon 
equation.

It is easy to show, with the operator expression of the SSE, that the free 
radial solutions are $R_l(r) \div j_l(kr)$ where the function $j_l(kr)$ is 
the regular spherical Bessel function defined as
\begin{equation}
\label{eq43}
j_l(x)=\sqrt{\frac{\pi}{2x}}\, J_{l+\frac{1}{2}}(x), 
\end{equation}
where the function $J_\nu(x)$ is the Bessel function of the first 
kind~\cite[p. 951]{grad80}. The wave number $k$ is given by
\begin{equation}
\label{new6}
k^2 =\left(\frac{1}{4E^2}\left(m_1^2-m_2^2+E^2\right)^2-m_1^2\right).
\end{equation}
In appendix~\ref{appenda} we show that,
\begin{equation}
\label{new7}
\int_{0}^{\infty} {\cal G}_l(mr,mr')\, \gamma r'\, j_l(\gamma r')\, dr'=
\frac{\pi}{\sqrt{\gamma^2 +m^2}}\, \gamma r\, j_l(\gamma r).
\end{equation}
Then one can verify that Eqs.~(\ref{eq20}), (\ref{new2}) and (\ref{eq27}) 
are true for a vanishing potential when $\gamma$ is replaced by $k$. In 
appendix~\ref{appendb}, we show that the situation is rather different if 
one consider the irregular spherical Bessel functions.

\section{Analytical results}
\label{sec:analsol}
In this section we present some analytical results concerning both the 
action of the square-root operator and solutions of the SSE. We first 
calculate the action of the kinetic energy operator on the functions 
$r^n\, \exp(-mr)$ ($n \geq -1$ and $n$ integer). The Fourier transform, 
$\tilde{f}(p)$, of these functions are
\begin{equation}
\label{new8}
\tilde{f}(p)=(-)^{n+1}\sqrt{\frac{2}{\pi}}\, \frac{\partial\,^{n+1}}
{\partial\, m^{n+1}}\, \frac{1}{p^2+m^2}.
\end{equation}
Then for each value of $n$ we can calculate the action of the square-root 
operator. We have, for $n=-1$ and $n=0$,
\begin{equation}
\label{new9}
\sqrt{\vec p\,^2+m^2}\, \frac{\exp(-mr)}{r}=\frac{2m}{\pi r} K_1(mr),
\end{equation}
\begin{equation}
\label{new9bis}
\sqrt{\vec p\,^2+m^2}\, \exp(-mr)=\frac{4m}{\pi} K_0(mr).
\end{equation}

We can get two more relation knowing the following 
integrals~\cite[p. 429]{grad80}:
\begin{equation}
\label{new10}
\int_{0}^{\infty} \frac{x^{2b+1}\, \sin(ax)}
{\left(m^2+x^2\right)^{n+\frac{1}{2}}}\, dx=(-)^{b+1}\frac{\sqrt{\pi}}
{2^n\, m^n\, \Gamma\left(n+\frac{1}{2}\right)} \frac{\partial\,^{2b+1}}
{\partial\, a^{2b+1}}\Bigl[a^n\, K_n(ma)\Bigr],
\end{equation}
with $a>0, \text{Re}\, m>0, -1\leq b\leq n$ and
\begin{equation}
\label{new11}
\int_{0}^{\infty} \frac{x^{2b+1}\, \sin(ax)}
{\left(m^2+x^2\right)^{n+1}}\, dx=(-)^{b+n}\frac{\pi}
{2\, n!} \left[\frac{\partial\, ^n}{\partial\, \gamma^n} \left(\gamma^b\, 
\exp(-a\sqrt{\gamma})\right)\right]_{\gamma=m^2},
\end{equation}
with $a>0, 0\leq b\leq n, |\text{arg}(m^2)|<\pi$. Rewriting these integrals 
as 3-dimensional integrals, extracting the square-root operator and 
integrating the resulting integrals one obtains the following formulas:
\begin{equation}
\label{new12}
\sqrt{\vec p\,^2+m^2}\left\{\frac{1}{r}\left[\frac{\partial\, ^n}
{\partial\, \gamma^n} \left(\gamma^b\, \exp(-r\sqrt{\gamma})
\right)\right]_{\gamma=m^2}\right\}=(-)^{n+1}\frac{2\, n!}
{\pi\, m^n\, (2n-1)!!\,r}\frac{\partial\,^{2b+1}}
{\partial\, r^{2b+1}}\Bigl[r^n\, K_n(mr)\Bigr],
\end{equation}
and
\begin{eqnarray}
\label{new13}
\nonumber
\sqrt{\vec p\,^2+m^2}\left\{\frac{1}{r}\frac{\partial\,^{2b+1}}
{\partial\, r^{2b+1}}\left(r^{n+1}\, K_{n+1}(mr)\right)\right\}&=& 
(-)^{n+1}\frac{\pi\, m^{n+1} (2n+1)!!}{2\, n! \, r} \\
&&\left[\frac{\partial\, ^n}{\partial\, \gamma^n} 
\left(\gamma^b\, \exp(-r\sqrt{\gamma})\right)\right]_{\gamma=m^2}.
\end{eqnarray}
These two equations are valid for $0\leq b\leq n$.

With these results we can find easily some analytical solutions of the SSE                   
with a nonlocal interaction. Let us consider the following equation,
\begin{equation}
\label{new14}
2\sqrt{\vec p\,^2+m^2}\, R_0(r)=E\, R_0(r)-\int_{0}^{\infty} W(r,r')\,
R_0(r')\, dr'.
\end{equation}
If we choose 
\begin{equation}
\label{new15}
W(r,r')=\left[\alpha\, \exp(-mr)-\frac{8m}{\pi}\, (\beta+m)\,
K_0(mr)\right]\, \exp(-\beta r'),
\end{equation}
then using Eq.~(\ref{new9bis}) we obtain the solution
\begin{eqnarray}
\label{new17}
&&R_0(r)\div \exp(-mr),
     \\
&&E=\frac{\alpha}{\beta+m}.
\end{eqnarray}
Thus it is easy to construct a set of solutions with the operator 
expression of the SSE, we must just consider a well chosen interaction. 
We can also find analytical solutions using Eq.~(\ref{new5}) with a 
nonlocal interaction: 
\begin{equation}
\label{new18}
\frac{2}{\pi} \int_{0}^{\infty}\tilde{{\cal G}}_l(mr,mr') \, u_l(r')\, dr'
=E\, u_l(r)-\int_{0}^{\infty}\, W(r,r')\, u_l(r')\, dr'.
\end{equation}
Choosing, for example, $l=0$ and
\begin{equation}
\label{new19}
W(r,r')=-\frac{2}{\pi}\, {\cal G}_0(mr,mr')\left[\frac{2\beta}{r'}+(m^2-
\beta^2)\right]+\gamma r\, \exp(-\beta r-\alpha r'),
\end{equation}
then we get the solution
\begin{eqnarray}
\label{new20}
&&u_0(r)\div r\, \exp(-\beta r) \\
&&E=\frac{\gamma}{(\alpha+\beta)^2}.
\end{eqnarray}
Again we can construct with this representation of the SSE a set of 
analytical solutions corresponding to a set of well chosen interactions.

\section{Summary}
\label{sec:summary}
In this work our purpose was to rewrite the spinless Salpeter 
equation into forms more easy to handled, since its differential operator 
is so particular. To perform it we have extracted the hidden nonlocality 
of the kinetic energy operator to get it really explicit. Thus we have 
obtained in Sec.~\ref{sec:theory} the following equations:
\begin{itemize}
\item For $m_1 = m_2$,
\begin{displaymath}
u_l(r) =\frac{1}{2\pi} \int_0^\infty {\cal G}_l(mr,mr')\, \Bigl(E - V(r')
\Bigr)\, u_l(r')\, dr', 
\end{displaymath}
and,
\begin{displaymath}
\left[\frac{d^2}{dr^2}-\frac{l(l+1)}{r^2}-m^2+\frac{E\Bigl(E-V(r)\Bigr)}
{4}\right]\, u_l(r) = \frac{1}{2\pi} \int_0^\infty 
\tilde{{\cal G}}_l(mr,mr')\, V(r')\, u_l(r')\, dr',
\end{displaymath}
with
\begin{displaymath}
\tilde{{\cal G}}_l(mr,mr')={\cal G}_l(mr,mr')\left[-\frac{d^2}{dr'^2}+
\frac{l(l+1)}{r'^2}+m^2\right],
\end{displaymath}
\item And for $m_1 \not= m_2$,
\begin{displaymath}
u_l(r)=\frac{1}{\pi \left(m_1^2-m_2^2\right)} \int_{0}^{\infty} 
\tilde{{\cal G}}_l(m_1r,m_1r')\, \Bigl(E-V(r')\Bigr)\, u_l(r')\, dr'
+(m_1 \rightarrow m_2).
\end{displaymath}
\end{itemize}
The kernel ${\cal G}_l(mr,mr')$ was first derived in Ref.~\cite{nick84}. 
It is analytic for each value of angular momentum and is given in 
Sec~\ref{sec:theory}. There are two integral equations and one 
integro-differential equation. These kind of equations are well known. The 
problem of definition, connected to the particular kinetic energy operator, 
are remove when the spinless Salpeter equation is rewritten into these 
forms or into the form~(\ref{new5}) (derived in Ref.~\cite{nick84}). 
Moreover with this last one we can calculate the action of the 
square-root operator on any functions of the form $h(r)\, Y_{lm}(\hat r)$ 
and find the radial part of the resulting functions. If $g(r)$ is this 
radial part, we have 
\begin{displaymath}
g(r)=\frac{1}{\pi r}\int_{0}^{\infty} \tilde{{\cal G}}_l(mr,mr')\, 
\tilde{h}(r')\, dr',
\end{displaymath}
with $\tilde{h}(x)=x\, h(x)$. When the integral in the right-hand side is 
relevant, this expression gives meaning to the action of the square-root 
operator on central problem wave functions. Indeed, this last relation 
allows us to explicitly calculate the resulting functions. 

In the Sec.~\ref{sec:analsol} we have found some analytical results 
concerning the action of the square-root operator on some particular 
functions. With these results and the integral formalism we have shown how 
to construct a set of analytical solutions of the SSE with well chosen 
interactions.

\acknowledgments
We thank Dr C. Semay and Prof. R. Ceuleneer for their availability and 
useful discussions.

\appendix
\section{}
\label{appenda}
In this section we show, with the integral equation formalism, that the 
free radial solutions are the regular spherical Bessel functions. It is 
easy to prove it with the operator expression of the SSE because one can 
rewrite it as
\begin{equation}
\label{apa1}
\left(\Delta+k^2\right)\, \Psi(\vec r\,)=0,
\end{equation}
where $k$ is given by Eq.~(\ref{new6}).

To prove it with the integral formulation, we use the integral expression 
of the kernel ${\cal G}_l(mr,mr')$ (see Eq.~(25) Ref.~\cite{nick84})
\begin{equation}
\label{apa2}
{\cal G}_l(mr,mr')=\frac{m^2}{2}\, r\, r'\, \int_{0}^{\infty}\, 
\exp\left[-\frac{1}{u}-\frac{m^2}{4}\left(r^2+r'^2\right)u\right]\, 
i_l\left(\frac{1}{2}m^2r r' u\right)\, du.
\end{equation} 
Then knowing that
\begin{equation}
\label{apa3}
j_l(x)=\sqrt{\frac{\pi}{2x}}\, J_{l+\frac{1}{2}}(x),\ 
i_l(x)=\sqrt{\frac{\pi}{2x}}\, I_{l+\frac{1}{2}}(x),
\end{equation}
and \cite[p. 718]{grad80}
\begin{equation}
\label{apa4}
\int_{0}^{\infty} x\, \exp(-\alpha x^2)\, I_{\nu}(\beta x)\, 
J_{\nu}(\gamma x)\, dx=\frac{1}{2\alpha}\, \exp\left(\frac{\beta^2-\gamma^2}
{4\alpha}\right)\, J_{\nu}\left(\frac{\beta \gamma}{2\alpha}\right),
\end{equation}
(with $\text{Re}\, \alpha > 0$, $\text{Re}\, \nu > -1$), we have 
\begin{eqnarray}
\label{apa5}
\int_{0}^{\infty}{\cal G}_l(mr,mr')\, \gamma r'\, j_l(\gamma r')\, dr'&=&
\frac{\sqrt{\pi}}{m}\, \gamma r\, j_l(\gamma r)\, 
\int_{0}^{\infty} \frac{\exp\left(-\frac{1}{u}\left(1+\frac{\gamma^2}{m^2}
\right)\right)}{u^{\frac{3}{2}}}\, du \nonumber \\
&=& \frac{\pi}{\sqrt{\gamma^2+m^2}}\, \gamma r\, j_l(\gamma r).
\end{eqnarray}

\section{}
\label{appendb}
We know that the relation
\begin{equation}
\label{ap1}
\sqrt{-\Delta+m^2}\, G(r)=\sqrt{\pm \beta^2+m^2}\, G(r),
\end{equation}
leads to $G(r)\div j_l(\beta r)$ (plus sign) or $G(r)\div i_l(\beta r)$ 
(minus sign). The situation is rather different for the irregular 
spherical Bessel functions $n_l(x)$ and $k_l(x)$ ($n_l(x)=\sqrt{\frac{\pi}
{2x}}N_{l+\frac{1}{2}}(x)$, where $N_{\nu}(x)$ is the Bessel function of 
the second kind). To illustrate, we just consider the function 
$k_0(x)=\exp(-x)/x$. We see that the right-hand side of the 
Eq.~(\ref{new9}) is different from zero as we could expect from 
Eq.~(\ref{ap1}). Then we can write
\begin{equation}
\label{ap4}
\sqrt{\vec p\, ^2+m^2}\ \frac{\exp(-\gamma r)}{r}=\sqrt{m^2-\gamma^2}\ 
\frac{\exp(-\gamma r)}{r}+f(r),
\end{equation}
where,
\begin{equation}
\label{ap5}
\left(\sqrt{\vec p\, ^2+m^2}+\sqrt{m^2-\gamma^2}\right)\ f(r)=4\pi \delta
(\vec r\,),
\end{equation}
since
\begin{equation}
\label{ap6}
\left(\Delta - \gamma^2\right)\ \frac{\exp(-\gamma r)}{r}=-4\pi 
\delta(\vec r\,).
\end{equation}
From Eq.~(\ref{ap5}) we deduce
\begin{equation}
\label{ap7}
f(r)=\frac{2}{\pi r} \int_0^\infty \frac{p\ \sin(pr)}{\sqrt{p^2+m^2}+
\sqrt{m^2-\gamma^2}}\, dp.
\end{equation}

Thus we see that the singularity at the origin of the functions $n_l(x)$ 
and $k_l(x)$ is very important and very annoying. Indeed, if $f(r)$ was 
null it would be easy to solve the equal masses relativistic coulomb 
problem for $l=0$. In this case one can show that
\begin{equation}
\label{ap9}
\sqrt{\vec p\, ^2+m^2}\ \exp(-\gamma r)=\sqrt{m^2-\gamma^2}\ 
\exp(-\gamma r)+\frac{\gamma}
{\sqrt{m^2-\gamma^2}}\ \frac{\exp(-\gamma r)}{r}
\end{equation}
and,
\begin{eqnarray}
\label{ap10}
\nonumber
\sqrt{\vec p\, ^2+m^2}\ r\exp(-\gamma r)&=&\sqrt{m^2-\gamma^2}\ 
r\exp(-\gamma r) +\frac{2\gamma}{\sqrt{m^2-\gamma^2}}\ \exp(-\gamma r)- \\ 
&&\frac{m^2}{\left(m^2-\gamma^2\right)^{\frac{3}{2}}}\ 
\frac{\exp(-\gamma r)}{r}.
\end{eqnarray}
Acting the square-root operator on the left of these relations one can 
verify them. 
The equation to solve is
\begin{equation}
\label{ap11}
2\, \sqrt{\vec p\, ^2+m^2}\ R_l(r)=\left(E+\frac{\kappa}{r}\right)\ R_l(r). 
\end{equation}
Using Eq.~(\ref{ap9}), we find that the ground state is 
\begin{equation}
\label{ap12}
R_0(r)\div \exp\left(-\frac{\kappa m}{\sqrt{4+\kappa^2}}\ r\right)
\end{equation}
with,
\begin{equation}
\label{ap13}
E=\frac{2m}{\sqrt{1+\left(\frac{\kappa}{2}\right)^2}}.
\end{equation}
And, using Eqs.~(\ref{ap9}) and (\ref{ap10}), the first excited state is 
given by
\begin{equation}
\label{ap14}
R_0(r)\div \left(1-\frac{16\kappa m}{\left(16+\kappa^2\right)^{\frac{3}
{2}}}\ r\right)\ \exp\left(-\frac{\kappa m}{\sqrt{16+\kappa^2}}\ r\right)
\end{equation}
with,
\begin{equation}
\label{ap15}
E=\frac{2m}{\sqrt{1+\left(\frac{\kappa}{4}\right)^2}}.
\end{equation}
This spectrum is a rather good approximation. For example, the error is 
$\approx \, 0.25\%$, for the ground state, if $\kappa = 0.456$ and $m=1$ 
GeV (see Ref.~\cite{brau97}). These energies were first obtained in  
Ref.~\cite{dura83}. Let us remark that the wave functions are 
Schr\"odinger-like wave functions.

\end{document}